# Aiming AI at a Moving Target: Health (or Disease)

Mihai Nadin[1]

Mihai Nadin
nadin@utdallas.edu
https://www.nadin.us

[1] Institute for Research in Anticipatory Systems
University of Texas at Dallas, Richardson TX, USA

**Abstract**: *AI and Society*—the encompassing subject of this journal—is probably illustrated (although not exhausted) by interrogations specifically focused on AI in medicine. Justified by spectacular achievements recently facilitated through applied deep learning methodology (based on neural networks), the "Everything is possible" view dominates this new hour in the "boom and bust" curve of AI performance history. The optimistic view collides head on with the "It's not possible," ascertainments often originating in a skewed understanding of both AI and medicine. The meaning of the conflicting views—whose score reality will eventually settle—can be assessed only by addressing the nature of medicine. Specifically, by answering the question: Which part of medicine, if any, can and should be entrusted to AI—now or at some moment in the future? That both medicine and AI will continue to change goes without saying.

**Keywords**: anticipation, algorithm, deep medicine, artificial intelligence, data types

Medicine is focused on what is needed to maintain life. It is an endeavor within the larger context of social organization of productive activity, of economic and political interaction, of culture. These and the environment, which used to be acknowledged as the dominant factor before the genetics revolution, affect medicine in all its aspects. AI in medicine is the shorthand for the meeting point between a new technology and what healthcare practitioners are or should be. Specifically:

- How to identify the specific talent and dedication healthcare demands
- How medical education should be conceived and carried out

- How to define experience—which medicine depends on—as well as blinding bias. (The "fresh eye" of a colleague or colleagues can help.)

While it is true that the beginnings of medicine are muddled—how much understanding was based on observation (empirical knowledge) vs. how much conjuring of the magical—the vector of development has been oriented towards ever more science and technology. Each step forward in knowledge acquisition and dissemination echoed in the practice of healing and maintaining a healthy life. Regardless of which new means and methods medical practitioners have adopted, major considerations of the larger context are kept in mind. Neither now nor in the past has the newest science and technology operated in a vacuum.

Within the framework of AI and medicine, access to care, the nature of treatment, the nature of work and leisure (as they affect health and healing), the roles of race and gender, of nutrition and exercise, and of other distinctions in providing and receiving care come into focus. So do the economics (who can pay for it?), the politics (who should get it?), and the culture of care. For such considerations alone, it would be wise to notice that AI, or any other perspective of science and technology-based medicine, would have to deal with an extremely complicated subject. Considering the history of medicine—almost as old as the history of the human species—an informed observer, i.e., somebody cognizant of the long-range outcome of medical care, would probably agree that medicine addresses the undecidable—and therefore intractable—nature of life's changes. Through the qualifier "undecidable"—as Gödel defined it in a different context—we understand the evasive nature of a subject that does not allow for a complete and consistent description (an issue addressed in detail in Nadin 2013). If AI—in whichever of its embodiments (e.g., symbolic, sub-symbolic, statistical) ever attains the ability to convert the undecidable into an effective tractable procedure, medical professionals, together with the rest of the world population, would be in luck. To be clear: The Utopian hope actually means that life itself would end since the living would return to its decidable non-living physico-chemical condition. We can repair bridges and keep machines in good running order. When it comes to the living (and this applies as well to animals and plants), the goal becomes elusive. And it will remain so because that is the nature of life. The fact that AI—which means those who are involved (as scientists, technologists, or investors) in a particular form of science and technology claiming credit for artificial intelligence—is interested in medicine has many explanations. The simplest (a bit cynical): Healthcare is the second most important sector of the USA economy (ca. 20%, which translates into trillions of dollars). Economic justification (or exploitation) aside, the challenges of taking care of more and more people, affected by more and more conditions that new forms of life and work entail, are real. Everything that can help—provided that no long-term consequences nefarious to life or the world result—should be considered.

## It all started with Shannon

There is a common denominator to the way medicine in what is called the information age changed from an empirical practice to a data-driven activity. To take note of how *artificial intelligence in medicine* (AIM), *deep medicine* (DM), *precision medicine* (PR), *digital medicine* (DMI), *high-performance medicine* (HPM), and *convergence medicine* (CM) reflect the focus on data is to gain access to the conceptual foundation of those currents.

To understand the condition of medicine in the information age, we need to understand not only what medicine is, but also what this information age, within which AI emerged, means. *Information Age*—the still fashionable branding of what defines the post-industrial age—is, in its own way, a recognition of Claude Shannon's contribution, and of a misunderstanding as relevant as what the term means to those building upon his legacy. His contribution: Information Theory. In fact, it is the mathematical foundation for the engineering of data transfer, over a given channel, from one point to another. The misunderstanding: Data are not information. In his own words, "…semantic aspects of communication are irrelevant to the engineering problem," (Shannon 1948, p. 379). The meaning of data is irrelevant to the transmission problem. Analogy: transporting wares from A to B; the container is agnostic of what it contains. And even more, of how it will be used. Or yet more significant, where and how the data originated. Indeed, data can result from observing reality (measurement) or from modeling or simulating reality. While transporting data is independent of how they are acquired or generated, the interpretation, which is not part of Shannon's mathematical theory, and the use are context dependent. From the perspective of medicine, which is, by necessity, the domain of meaning (of changes in a person's condition), Shannon's work on how data are quantified and transmitted is significant only in regard to how data become available to the physician. More precisely, if it is actionable, in the sense of guiding interventions, such as surgery or the use of pharmaceuticals, not to mention genetic processes of healing.

Highlighting Shannon's contributions as the forerunner of the computing revolution and of AI is relevant for understanding what these actually mean. Before the computer, Shannon conceived a mathematical theory of circuits based on Boole's binary logic (Boole 1854) of Yes and No (eventually represented by 1 and 0). Shannon's were not the semiconductor-based circuits (the transistor had not yet been invented), but relays and switches: the electromechanical hardware, upon whose basis telephony (connecting households offices, and eventually cities, etc.) was developed. The mouse in the electro-mechanical maze he built was learning before the knowledge domain of Machine Learning was advanced. Project “Theseus” (named after Theseus of Greek mythology, who navigated the Minotaur’s labyrinth following a thread) is paradigmatic of what AI, in both its logic-based and statistics-driven phase would become. Copper whiskers were not yet the sensors of the generation after Shannon, but they indicate a direction of development relevant to deep learning methods deployed in cognitive studies, or in assessing motoric expression.

“Any circuit is represented by a set of equations,” which Shannon showed “is exactly analogous to the calculus of proportions used in the symbolic study of logic,” (1938). The definition encapsulates the expert systems phase of AI. It is important to notice that logic gates, which are what the computer chip integrates, began as electromechanical contraptions “trained” to learn arithmetic. Associated with this was the focus on quantifying data, i.e., measuring data. John Tuckey, the Princeton University professor in residence at the Bell Lab where Shannon worked, came up with the word “bit” when Shannon asked for a good name to describe binary choices. Shannon, however, made the bit describing the quantity of data corresponding to a Yes or No situation the elementary unit of a measuring system that everyone, doctors included, have taken for granted.

Evidently, when the gates were printed into integrated chips, processing became faster. The amount of data that could be processed increased exponentially. Today we are contemplating the zettabyte and the yottabyte, and even the brontobyte and the gegobyte (10 to the power of 30). However, not even these are large enough if we pursue the questionable path of monitoring every human being, from cradle to grave. Shannon never imagined such scale of data, and even less what it takes to transmit it effectively when a vast variety of data types are generated. The inescapable limitation of the mathematics describing the transmission process, i.e., the implicit characteristic of any data processing, is that it is void of meaning. This limitation is characteristic also of the Turing machine, which mathematics already discovered can be automated using exactly what Shannon described in his theory of circuits. As we shall see, there is no automatic procedure for creating mathematics (the *Enstscheidungsproblem* challenge, Hilbert, Ackerman 1928), because meaning—the outcome of interaction—cannot be generated by any kind of machine.

## Hoping for the human-artificial convergence

AI is part of the new computer-centric culture within which a new human condition is being shaped. Rich documentary evidence maps the influence of all tools—from the lever to the hydraulic, pneumatic, and electric machines that precede the computer—on those who conceived and used them. The cognitive and motoric profile of the human being in the age of computation changes before our eyes, affected as they are by the amount of time in front of the monitor and by the nature of interactions with machines. And sometimes not for the better. It is less than an exaggeration to state that to help cure some of AI's possible consequences on health (e.g., mental, emotional, physical) would be worth celebrating, just as much as recognizing what AI has contributed to improving medical care. Academics from respected institutions of higher education have already raised questions about the post-human centric civilization. Within the current understanding of what AI is, a theme that percolates is the replacement of human beings on factory floors, at their car's steering wheel, in farming, in financial services, and even in lawyering. And—why not?—in medicine. A reputable medical journal considers the "Perspective" (the column title under which it was published) of "Self-Facilitated Service in Healthcare” as a subject worthy of discussion (Asch, Nicholson, Berger 2019). Doctors debate whether AI will replace them, or make them obsolete (Goldhahn et al 2018, Budd 2019), or why this cannot happen (Medical Futurist 2018). Moreover, one of the famous Silicon Valley investors (Vinod Khosla) is on record stating that robots will replace doctors by 2035 (Kocher & Emmanuel 2019). A distinguished medical practitioner goes beyond peddling all kinds of devices and affirms the new stage of *Deep Medicine: How Artificial Intelligence Can Make Healthcare Human Again* (Topol 2019a). Before the ink on the book’s cover dried, the same author went for “High-performance medicine: the convergence of human and artificial intelligence” (Topol 2019b), leaving open the question of whether the relationship between doctor and patient will improve or erode further.

None of these subjects—bait for headlines—can be ignored. And none of the authors deserves less than attention to their ideas. It is honorable to serve as advisors to the many large and small attempts to reinvent the oldest known knowledge domain that healing and preserving life are. By the same token, such efforts should not escape scrutiny—no conflict of interest--, given the fact that in all cases of interest to medicine, the focus is on nothing else but life.

**A bit of history**

The history of medical care (not to be rehashed here) reflects the never-resolved question of what medicine is. (Medicine eventually extended to animal and plant diseases.) As a practical experience, it came close to witchcraft, as it was (and at times remains) close to religion—if for no other reason than it deals with life and death, subjects that can conjure belief more than understanding. Offerings to gods and goddesses of all kinds did not exclude parallel rational

paths—often against religion-based interdictions, such as the prohibition against dissecting cadavers, or against certain types of surgery. If spells did not cure, and amulets did not prevent disease, plants and minerals were used in a series of procedures that continue to our days. Artifacts identified from a present-day perspective as precursors to the scalpels that surgeons used, or as saws, scissors, even probes, document how far back humans adopted technology in their attempts to fix what affected the well-being of those suffering from pain and disease, or experiencing unusual emotions. Well over 2000 years ago, cataracts were removed, gangrened limbs were amputated, teeth were pulled, bones were set, herbal remedies were applied, prostheses were attached. Humans learned about their own make-up: to embalm a cadaver, i.e., to preserve it ("for life in the world beyond the grave") entailed removal from the body of a number of organs subject to degradation. Not feeling well or experiencing some breakdown was looked at not only through the eyes of physics and chemistry. Yes, most of the time, healers literally looked for what might cause pain, or what might alleviate someone's condition. The right balance of hot and cold, of wet and dry, of humors, of blood (treating inflammation and fever through bloodletting using leeches) reflect a naive understanding of the material reality expressed in the language of philosophy, and in early texts that predate physics, chemistry, and biology. Scholars dedicated to the pre-history of medicine are unanimous in regarding how tools influenced their own health and preoccupation with its possible loss. Conceived and utilized in the self-constitution of human beings through what they did in order to survive and better their chances of survival (Nadin 1991), such tools find their way into the hands of healers. This continues down through our times.

The understanding of reality has always been reflected in in the understanding of health and disease, as well as in all known forms of identifying whatever affected them. Each culture—of Babylon, China, and India, of Greece and the Roman Empire, of the Middle Ages (e.g., Rogerius's *Practice of Surgery*, ca. 1180), of the Renaissance (the microscope can be indexed to it as the beginning of systematic medical training)—marked the practice of healing through views and methods in accord with that culture's values. The infatuation with machines pursuant to the Cartesian Revolution brought about not only a different understanding of how compromised health affected performance of the "human-machine," but also a propensity towards deploying means and methods for fixing what "broke down." Just for the sake of illustrating this thought: in the Middle East, irrigation based on rudimentary hydraulic pumps inspired healers to seek where passages in the human body might become blocked (as happened with improperly attended irrigation canals), and why. A prayer, still recited daily by Orthodox Jews, brings up the body's "hydraulics" and expresses awe at the miracle of their functioning (if one cavity or opening is blocked, "we could not stand before You for even one hour;" *Talmud*, Berachot 60b). Egyptians, alert to the dangers of intestinal malfunctioning came up with what today we call laxatives (to ease the "blockage"), and with substances that triggered vomiting (Bryan 1932). As time passed, the model of the hydraulic machine was replaced by that of the pneumatic engine, and then by the combustion motor, by electric engines, electric circuits and relays, by the production line, and all that followed. The human being became the "machine du jour," not only as a description of how well it functions (when in a healthy state), but also in relation to how it can be fixed (when the "hardware" or "software" fails). In the civilization landing in the Turing machine universe (yet another cognitive horizon), everything becomes a matter of computation (with consequences for hardware and software design and implementation, for computational algorithms and for digital representations).

**Humans are neither computers nor neural networks**

To reject the formulation of a Turing Award laureate (the Turing Award is self-stylized as the Nobel Prize in computation) is not a task that an author endeavors without some hesitation. Science (and more so medicine) frowns upon challenging authority. But since the consequences of adopting a certain perspective in medicine are of long, and very long impact, it would be at least questionable not to point to what could become extremely detrimental. Johnathan Zittrain (of the Berkman Klein Center for Internet and Society, Harvard University) asks: What if AI in healthcare is the next asbestos? (2019) Not long ago, lobotomy was considered a safe treatment for epilepsy.

This is not the first time that a certain form of knowledge is hypostatized; that is, the description of a certain aspect of reality is assumed to be as real as the words or images describing it. For example, in order to practice agriculture in desert-like conditions, irrigation systems were developed; when water paths got blocked, they had to be opened. Well, maybe the body was like such an irrigation system. Or even—the hypostatized phase—maybe humans are an irrigation system. Within this manner of thinking, descriptions of phenomena are identified with what is described. Constructs such as rituals, religion, myth, or the successive metaphors used by science document how this confusion was legitimized. Therefore, to quote a scientist, who became the “Godfather of Deep Learning” (i.e., Geoffrey Hinton) is less an exercise in pointing to self-deception and more an illustration of the thought: “We humans are neural networks. What we do machines can do” (Hinton 2019). Of course, we are not neural networks, just as we humans were never irrigation systems, or hydraulic pumps, or engines, or machines, or production lines.

There is a lot to learn from the dialectics of hypostatized knowledge. Religion contributed much to culture: the premise—a world created by some divinity—did not prevent human performance from reaching impressive accomplishments on account of belief. This holds true for the arts, as well as for the sciences (think about Newton). Or, to use another example: the metaphor of the human being as a machine—probably accepted by the vast majority of those who are active in medicine—inspired means and methods that have not only saved lives, but have also helped a huge number of people to continue a productive existence. However, the implications of hypostatized knowledge can be meaningfully discussed only after medicine itself is defined. This becomes even more clear when the "Deep Medicine" model of extensive data processing—pushed by those who are convinced that this is the medicine of the future—is put under scrutiny. Promoted from within medicine, this model argues for more data and more machines if the doctors want to find time for their patients, and even regain empathy after the so-called precision medicine has transformed those doctors into data-processing agents.

It is noteworthy that each previous adoption of descriptions of reality pertinent to physics, chemistry, and biology were consequential for understanding the human being and its change over time. The doctrine of the four humors (blood, yellow bile, black bile, phlegm) associated with Hippocrates, originated in ancient Mesopotamia and India, but extends to the modern terminology of hormones and antibodies. The link between humors and elements in the universe and atmospheric conditions is by no means to be discarded. The role of the environment in which humans live—air and water quality, noise, pollution, etc.—is still only marginally acknowledged. As is known, the germ theory of disease was more effective than the doctrine of humors. It explained the spread of diseases and also brought about effective preventive and treatment methods. Sanitation, by now expressed as a prerequisite for public health, and urban growth meant more than the physician's obligation to wash hands before surgery and before examining and treating a patient. It is also more than the standards associated with clinics and hospitals (the "repair shops" that mushroomed in the early 20th century, before becoming the profit mills they are in our days), with nursing homes and assisted living facilities.

## Medicine is about ethos

Regardless of the successive ways in which the physician's activity was carried out, and regardless of the means and methods, medicine is fundamentally focused on health and what happens when health deteriorates for some reason, known or unknown. The standard definition includes what is necessary to encourage healthy living, to avoid deterioration of well-being, to promote life and its continuous renewal. Such goals are more a desired outcome for whose attainment physicians examine people seeking their advice. Diagnosis—interpretation of symptoms (which are the "signs" of a medical condition)—is the concrete outcome of evaluating the many factors involved in defining someone's condition. From diagnosis to prognosis—which changes can be expected or predicted—and finally to some intervention, there is a continuum within which physician and patient interact.

No matter which metaphor was adopted—the hydraulic model of the heart as pump, the body as an assembly line, conveyor belt, or as a factory, the eyes as lenses, the brain as computer, etc.—what has justified the transfer of knowledge, from physics mainly, or from chemistry, is the hope that the science behind such descriptions could help fix an undesired state of a living being. In other words, how the knowledge underlying human activity (farming, handicraft, industrial production, automation, etc.) can be generalized to the being itself.

Paradoxically, it is not from health—the normal condition of the living—that medicine seeks answers, but rather from the way science treats what is broken down, what no longer functions as expected. Failed irrigation, low pump performance, assembly-line bottlenecks, scratched lens, programming errors—there is an answer to how these can be fixed, or replaced. If one compares the standard definition of "diagnostic" (in any dictionary or encyclopedia) and the definition of a medical diagnostic, one common thought comes into focus: inference based on symptoms. However, to infer from a physical phenomenon is different from inferences pertaining to the living.

The physician of our time, not unlike the engineer, is educated within the cause-and-effect perspective. Pathogenesis—the biological source of someone's condition—is the starting point for those trying to help patients. The word brings up emotions. *Pathos*, in Aristotle's view (Kennedy 1991, p. 119) would mean to understand the emotions: name them, describe them, figure out what might cause them. Indeed, human beings are emotional about anything affecting their health. Again Aristotle: to address someone's feelings (from frustration to fear), one has to act with *ethos*, i.e., credibility. That's why the history—anamnesis, i.e., the patient's account of the breakdown, as well as medical record—in combination with symptoms and examination (the current state) guide those preliminary assessments of aspects of a person's condition to be further investigated (tests, performance evaluation, etc.). Etiology—i.e., what might have caused a certain condition—goes hand-in-hand with the evaluation of symptoms. This is the beginning of a path that can be short—temporary conditions (transitory breakdowns) easy to address, such as a sore throat or earache, or minor accidents such as a gash in the skin, a swollen ankle—or tortuous (long-term targeted therapy in cancer), breakdowns that are systemic in nature.

Given the nature of complexity characteristic of the living—i.e., *G-complexity*, which means that the living cannot be completely and consistently described (as mentioned above)—neither the assessment of symptoms nor the formulation of etiology can be univocally determined. A physician's description is more precise than the expression of emotions, or the self-diagnosis provided by a patient (or anyone else), because a less ambiguous language is at work. In fact, the various specialized languages that doctors use are representations conceived with the purpose of capturing details. Differential diagnosis means that from a subset of the very large possible set of explanations—those apparently more probable—was identified. Physicians are trained, and experience further teaches them to pursue the task of identifying what is significant. From among the many possible explanations of the pathogenesis that make sense, only some are pursued. The same holds true for the etiology: compare potential answers to what might have caused a condition (possible causes) and seek correlations. Until recently, the microbiome (the bacteria, fungi, viruses, etc. that live inside the body) was not even considered. From a logical perspective, the open-ended induction (what big data tries to capture)—consider everything—is subjected to pruning: consider what makes sense in a given context. Which means: from big data to small, but *significant*, data. This is the classic preliminary detective work that rules out (not rarely, mistakenly) causes that make no sense in the context of the medical exam. The headache that goes away when you are in the doctor's office is obviously connected to a context different from that of the examination.

## Health is about the future

Of course, past and present (describing a medical condition) are not enough: the real subject is the future—which medicine still ignores (Nadin 2016, 1-27). Physicians know that symptoms are most of the time non-specific. Sweating might suggest anything from overheating (and wearing the wrong clothing) to cancer. Erythema, which means red skin color (but which red? Intense? Pinkish?) can be, in principle, connected to various etiologies. Add ambiguity to this: symptoms and their interpretations are not in a clear-cut relation. Various healthcare professionals (internists, surgeons, dentists, optometrist, etc.) proceed in a manner specific to their particular perspective, i.e., the language of their specialty. Blood test panels or a variety of imaging procedures (ultrasound, radiology, magnetic resonance, etc.) return to a diagnosing medical professional more detailed data acquired in a reductive manner: ignore the whole, focus on details. For the limited perspective they define—e.g., a specific liver or kidney function, the cartilage in the painful knee—the data acquired might suggest a certain intervention to alleviate some undesired symptoms, but rarely, if ever, to respond to the etiology: the "Why?" question. Unless this question is addressed, the future is reduced to the interval between two visits at the doctor's office.

The decision to delve more extensively instead of more deeply is the result of medicine's implicit rejection of the fundamental understanding of life as different from the non-living. *Deep Medicine* should have already acknowledged that depth—in awareness of the holistic nature of life processes—is the opposite of what "deep learning" (the technology of super-Big Data) can achieve. Indeed, the non-living can be completely and consistently described: statistics describe regularities. It is in the domain of the decidable. All physical phenomena described in the language of the mathematics of change (calculus) take place under the threshold of G-complexity. All phenomena of what we call *life,* at all levels of the living, take place over the threshold of G-complexity. Change in the non-living is described through a limited number of variables that remain the same within the process. Change in the living, including what is called symptomatic change, is, of course, also subject to causality. More precisely: multicausality. The multicausality characteristic of the living is different in nature from the physico-chemical causality associated with the fundamental physical forces (gravitational, weak nuclear, electromagnetic, strong nuclear) supposed to explain reality. On account of the current model (in progress) of the fundamental physical forces, we can explain gravity, as well as why our hands can't simply reach through a wall or a sheet of metal, how storms develop, how earthquakes form. But even people in full command of the mathematics describing such forces could not explain the pathogenesis of mental illness, or the creative outcome of aesthetic endeavors. The mechanical piano, or its digital embodiment, is not really a substitute for a live concert number, not to say for the sublime of a rendition by Vladimir Horowitz or, in our time, by Vladimir Ashkenazi. The music is not in the piano, as N.A. Bernstein, assisted by Tatiana Popova (1930) documented through their experiments on motoric expression in piano playing. For any living process the number of variables—i.e., changing details—is non-determinate. In this respect, the awareness of the role of the microbiome made this realization even more evident.

These variables define what in system science is called the *phase space*. This phase space describes the dynamics of the living system understood as a whole. The phase space of physical processes (e.g., falling of a stone, erosion of a beach, lightening) is constant: the number of variables needed to describe the process remains the same. The phase space of living processes, i.e., the number of variables necessary to describe their change, also changes (Longo 2018). The living is in a continuous process of making and remaking itself. This represents a real challenge for those who are trying to describe it—through measurement, visual representations, sound, in mathematical terms, or in computational models.

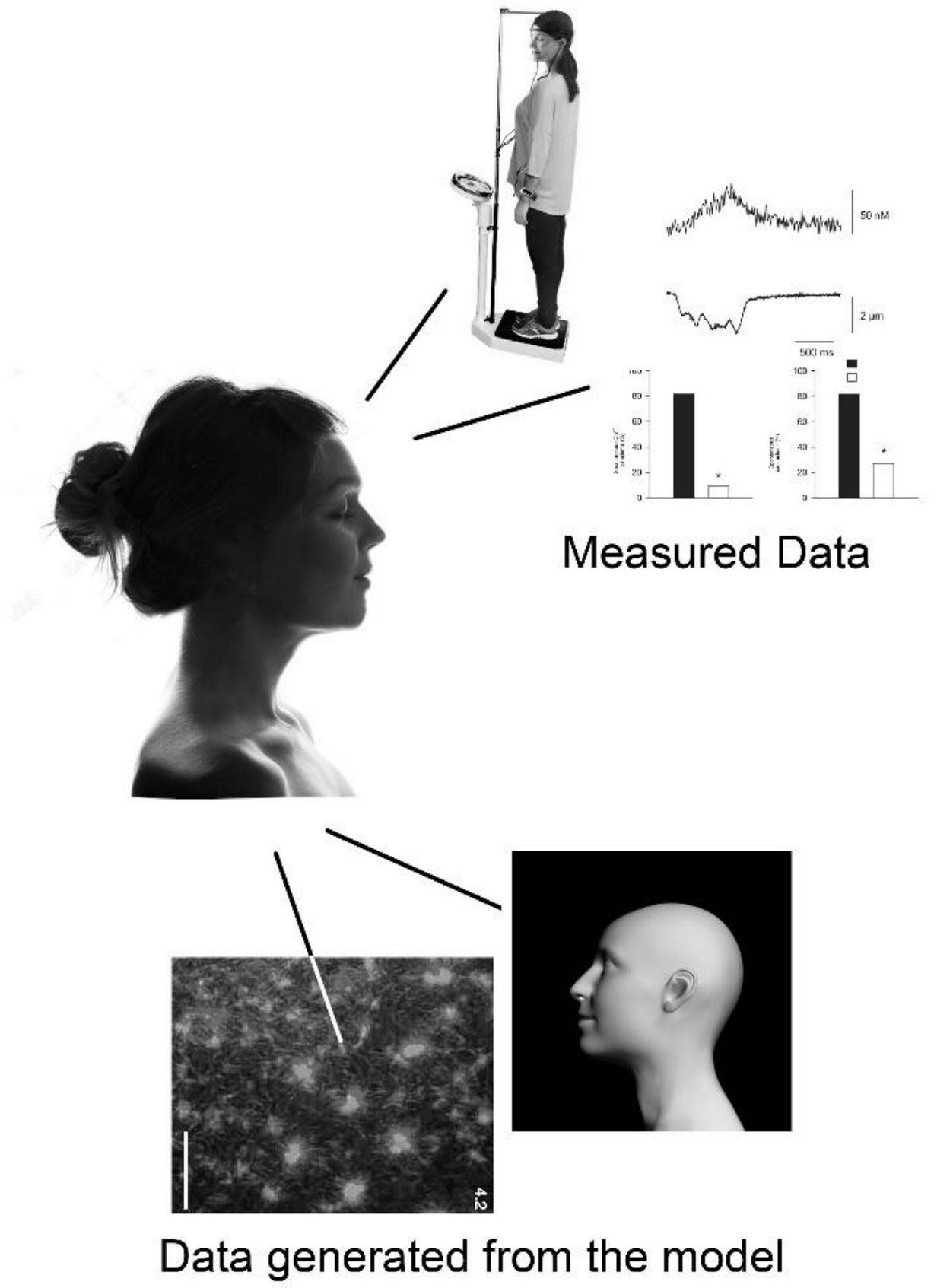

Fig. 1. Measured data (that can be used to train a neural network) vs. artificially generated data

The data, such as sensor data (representation of heart rhythm, or temperature, or saliva acidity, for example), or visualization, biochemical analysis (such as the blood test) have to be referenced to meaning; that is, the data has to be interpreted. Data generated from computational models (of a disease, for instance) or simulations (of a treatment) make sense only when referenced to the science behind the processes for generating them. Erythema, with red skin as a symptom, has a specific meaning in the context of an individual's life. Hot water accidentally spilled over one's hand is the answer to the "Why?" question. Targeted therapy, appropriate to what has to be treated, sometimes results in a prescription, other times in a course of treatment.

The difficulties in dealing with the undecidable explain why medicine, despite spectacular advances deserving more celebration than they get, remains extremely challenging. The territory to be mapped—a person's condition—continuously changes, often because the act of mapping—taking measurements—affects the landscape. If you've ever checked your blood pressure using the classic cuff, you will notice that simply watching the monitor affects the result of the measurement. Not that the gaze moves the dials, rather that feedback processes affect blood circulation, i.e., the heartbeat. Under a physician's observation, a patient undergoes change.

## What is life? A question that will never go away

Robert Rosen, the mathematician of biology who dedicated himself to defining what life is (1991), advanced the <M,R> model, i.e., a description that accounts for the empirical observation of metabolism and repair (self-repair, actually) as intrinsic to life. In his words, life is closed to the effective cause, that is, it takes place inside, not as a result of outside causes. There is genius to this description, which distinguished scholars have discussed in some detail, noticing, correctly, that it is not computable, i.e., it cannot be reduced to an algorithm (Louie 2013). On account of metabolism (the unity of anabolism and catabolism), life is maintained. This means that living matter (the elements from which it is made) is synthesized, but also decomposed (eventually) through biological processes. The energy, as well as the biological matter are acquired through interactions with the world in which life unfolds. The whole is hierarchically structured. Cells, the elementary units of life, are made and remade, ensuring the stability of the organisms against the background of incessant change. Repair is dependent on the metabolism—it requires energy and various ingredients, some self-generated, some gained through interactions with the environment. But the model is incomplete: the living undergoes, in addition to metabolism and self-repair, a permanent renewal of itself. As pointed out (Nadin 2019), all cells making up the body are renewed—at different rates of renewal, some more frequent than others, depending on the functions they perform within its life cycle. Moreover, of a higher level of necessity than metabolism and self-repair (which is different from renewal) is the creative nature of life: preservation of life is as fundamental as any, if not all, laws of physics (Nadin 2015).

Diagnosis in the non-living realm—e.g., the car engine, the lock, the window-opening mechanism—means description of its limited dynamics. For example: how the elements turn stone into sand over centuries; how weather changes; what earthquakes, or volcanic eruptions, are. Actually, everything, including the decomposition of dead humans, animals, trees, and other vegetation. In the particular case of artificial non-living entities—from new materials to machines and associated processes—such descriptions of dynamics constitute the basis for prognosis: the bridge might (or will) break down; the airplane design (does the Boeing 737 MAX come to mind?) is faulty (or not); a certain material is safe, poses no harm to humans. To describe behavior over time is, in such cases, to predict based upon descriptions that qualify as laws of physics. Everything in reality that appears as subject to laws (usually descriptions in the language of mathematics or logic, which means also the language of algorithmic computation as a form of automated mathematics) belongs to a domain identified by Windelband (1894) as *nomothetic*. Within the nomothetic, the cause-and-effect vector is well defined: from past to present. But health is also a matter of future. Take only aging as an example in order to understand what this means.

The living is no less part of the nomothetic domain, as everything else embodied in matter. Nevertheless, unavoidable nomothetic aspects are only a subset of the broader description that characterizes the uniqueness of each embodied individual. Not even twins are 100% identical. As for the rest of the living, no two cells, not to say no two blades of grass, are identical. To diagnose the living (human being, animal, plant) implies the awareness that from a very large phase space (large number of parameters), the physician will select a subset. Moreover, it implies the awareness that this subset itself is subject to change. Just for the sake of illustrating the thought: Within the 24-hour cycle of the day—not to mention the longer cycle of the seasons, nor the influences of sun and moon—some intervals are better suited for surgery or for other targeted therapies. Reducing the undecidable, intrinsic in the description of the living, to a decidable description is what a healthcare professional ultimately does (Fig. 1).

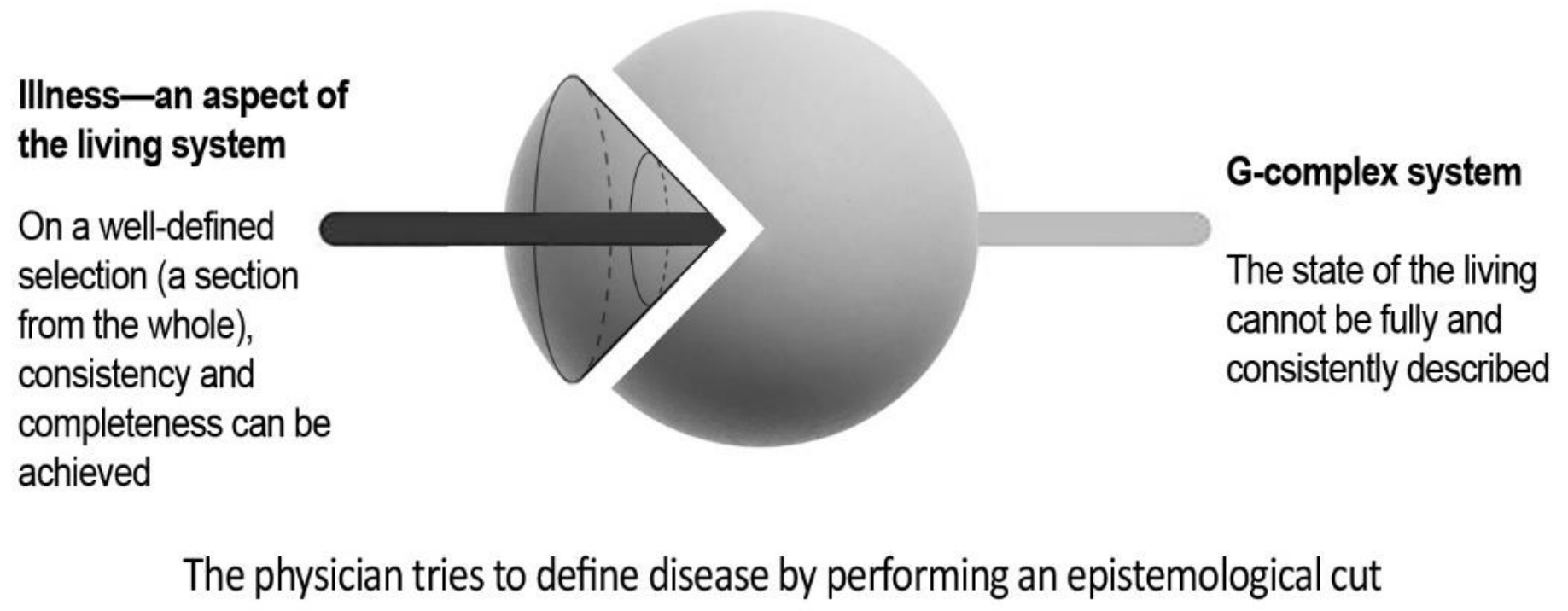

Fig. 2. The physician tries to define disease by performing an epistemological cut.

To diagnose a machine is to establish which parameter (or parameters) undermine the “law” embodied in the device. Data describe each and every deviation, and thus informs the action of repair. It does not matter if you replace a gear during the autumn, in the evening or early morning. Moreover, the number of parameters describing the machine’s functioning or a bridge’s characteristics does not fluctuate. (Their value can change, of course, as temperature, humidity, position, etc. vary.)

To diagnose the living is to understand its uniqueness. Data are necessary, of course. A broken arm bone is diagnosed only after a medical professional establishes the pathogenesis of the injury: tenderness, swelling, an identifiable open wound. The patient is engaged in the conversation; emotions are "translated": Is the pain strong or slight?  Is the patient under medication (self-prescribed, or from some legitimate or illegitimate source). Is the patient depressed or excited? To define the physics of the injury—greenstick fracture (the bone is not broken all the way through), a buckle fracture (bending of one side of the bone under the pressure of the other side), comminuted fracture (the bone is fragmented), etc.—the physician will first locate it either through an X-ray or a computerized tomography image. The effort is towards "encircling," delimiting the significant variables. Once the targeted actions for getting the process under control and eventually defining a therapy become the focus, the physician expands from the localized aspects: patient condition (age, heart condition, physiological aspects, nutrition, drug use, alcohol consumption, etc.). In other words, the phase space to be considered is different for the general practitioner, for the X-ray technician who takes the image and for the radiologist who will interpret it, for the surgeon, for the anesthetist, for the nurse, for the physical therapist.

To repeat: data are necessary. But not sufficient. Each broken bone comes with its history: a physical event (e.g., accident, fall) or a biological stage (age-related fractures, depletion of minerals). Data have to be referenced to the context in order to become meaning. Medicine is not a data-processing endeavor, but an art-and-science interpretation focused on meaning. For the nomothetic (i.e., anything not alive, not subject to physical law only), the more data—i.e., the more measurements performed—the better. For the *idiographic,* (i.e., anything unique, which the living is), *significant* data are what make diagnosis possible and meaningful. Moreover, chronobiology made us all aware of the role that time plays, not only in deciding when a surgery should be performed or the time when a medication is more effective. A study of the “emerging link between cancer, metabolism, and circadian rhythms” opens up an even wider window into the role of timing in respect to health and disease (Masri, Sassone-Corsi 2018). Computational model data, generated on account of a new hypothesis (a new view of disease, or a new method of treatment) are also necessary. To arrive at their meaning is probably even more difficult than to evaluate the meaning of data from observations of the real patient.

These considerations, repeating arguments made elsewhere (Nadin 2018a) will constitute the background against which understanding of what is going on in the medicine of the information age, in particular when AI is utilized, becomes possible.

## Medical data make sense in context

One first observation: when diagnosis is confused with the technologies deployed for harvesting data, the possibility to understand the “Why?” of life (and disease) is undermined. An X-ray of a non-living entity—a stone, an engine, a sculpture or painting—facilitates access to data describing the nature of the matter it is made of and its components (also known as parts). The X-ray does not alter the object (if it does, the alteration is minimal). Think about X-rays of paintings, performed in order to make visible what the naked eye cannot perceive. Or of X-rays of crystals. We get an image of the invisible structure, which is not affected (or minimally, at most). The X-ray of a breast—the mammogram prescribed for women after a certain age, regardless of their specific condition (context)—could help identify cancerous developments. But science already warned that the procedure's fail rate is a bit discouraging.

Strictly for illustrating the thought (and in preparation for the evaluation of AI-based medicine), here are some simple facts:

- In a false negative mammogram, everything looks normal, even when there is breast cancer
- At a 20% rate of error (one in five patients!), the data can be considered only under advice
- In a false positive mammogram, an image of a tumor appears, even when there are no cancerous cells in the breast

In addition, the phase space to be considered is defined by considerations that transcend the focus on breast cancer. Liver, kidney, lung, or heart disease can be more life threatening even in the case of breast malignancies. Only some will develop into conditions requiring treatment.

Worse yet: the attempt to measure is more disruptive in the living—radiation exposure, for example—than the attempt to measure the non-living. A physical object is minimally affected by the X-rays used to make visible internal structures (of crystals, machines, suitcases subjected to security checks at airports, for example). Such considerations, among others, explain why medicine should not blindly follow the path of collecting more data. Instead of more meaningless data, the option should be *less but significant*. In other words: what makes sense. Medical competence, not data processing skill, should guide the process of defining what is meaningful.

Echocardiography, the complete blood count (CBC), colonoscopy, magnetic resonance imaging (MRI), computer axial tomography (CT or CAT-scan)—very respectable technologies, often deployed in order to save lives, assist in datamining the body. The diagnosis—the outcome of patient-physician interaction—leads to defining targeted therapies. Nevertheless, the purpose of the process is to facilitate healing, not data collection. What makes this observation even more critical in the evaluation of what AI can contribute is the fact that each technology means a new language. The data do not “speak” the language of the physician, anesthetist, radiologist, or surgeon. Somebody has to translate. MRI images are not photographs, but computer graphic renditions. They are generated data based on the science of radio wave propagation. To interpret an MRI image presumes a different knowledge set than that required by the interpretation of an X-ray or CT scan. As we shall see, the AI of our time recognized an opportunity in automating data interpretation, even promising to replace the live human interpreter, or at least to become an indispensable assistant.

The timeline of medical events and initiatives that reflects the cognitive and epistemological condition of healthcare includes, as we have seen, those beginnings (harking back 5000-6000 years) of awareness of well-being in contrast to some breakdown. This awareness shadows the entire history of the human being. But only after the medical profession acquired its status in society can we speak of a body of specific skills and specific knowledge that differentiates the work of those who are focused on healing from other forms of work. Some milestones are to be acknowledged: understanding metabolism in a non-reactive manner informed the use of insulin for diabetes; understanding aspects of the immune system and its anticipatory nature led to the practice of vaccination (against diphtheria, pertussis, tetanus, tuberculosis, polio, yellow fever, typhus, influenza, etc.). Following the description of DNA structure, medicine adopted genetic methods, leading to the mapping of the human genome, and to attempts to manipulate the “program” for healing purposes. Antibiotics, a biochemistry-based means for fighting infections, mark a new direction, which eventually proved, due to overuse to have less than desired consequences.

Having entertained the question “What is life?” we realize that all new means and methods (from the history of medicine) that are based on the idea that the living are nothing other than a physical entity are translated into a technology of fixing (Rosen 1968). Here are some of the perspectives that have been advanced over time: the spare-parts view that knee and hip replacements embody (Fig. 2).

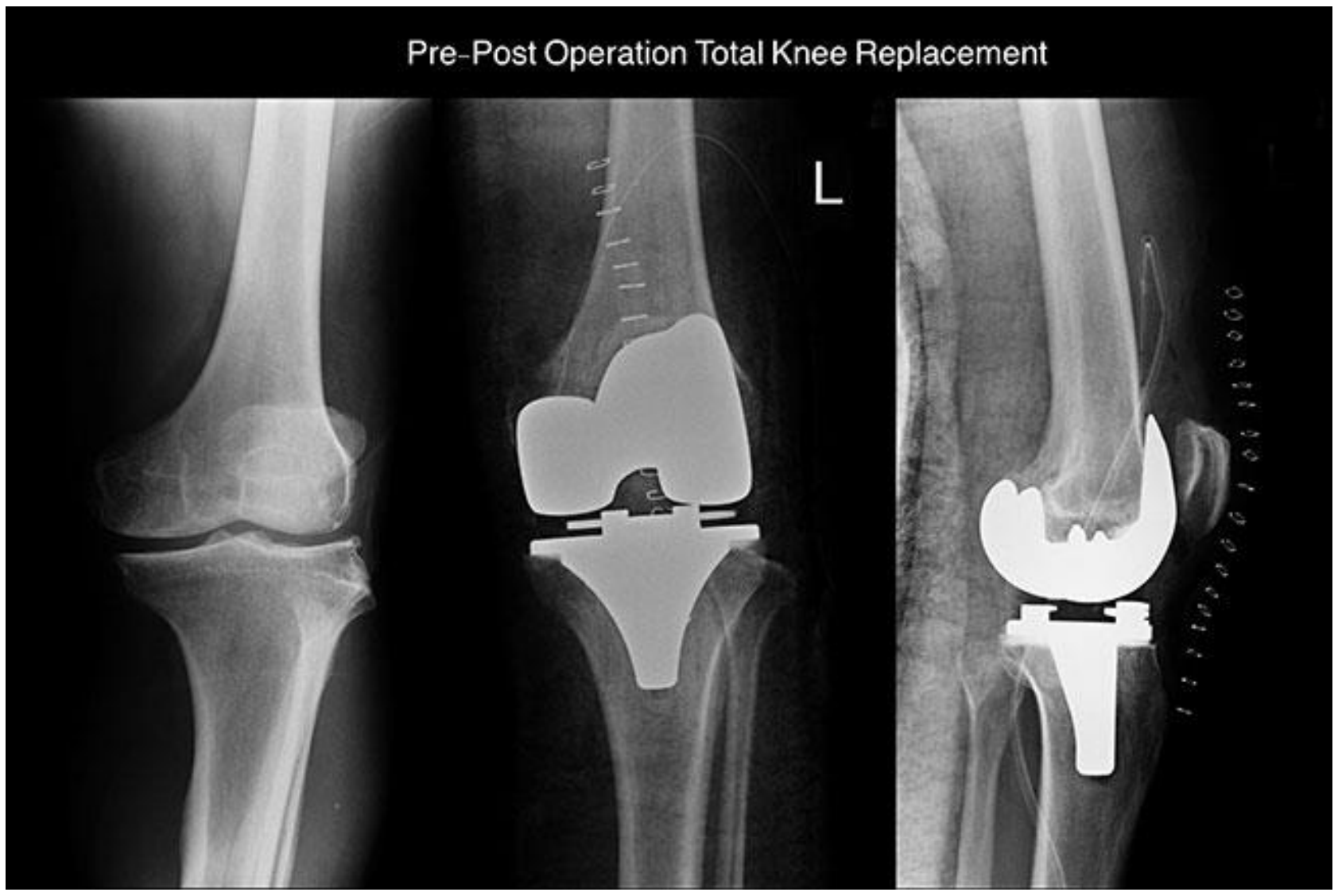

Fig. 3. The prosthesis exemplifies the “spare part” perspective of medicine as well as of mechanics. An alternative is genetic healing (self-repair).

In the same category: the view that life is a chain of chemical reactions (one in six Americans takes a psychotropic drug to address what is described as chemical imbalance). Or the view that life is a psychological subject, which translates into psychiatric sessions for those addressing the pathogeny of mental illness. The view that life is genetic expression leads to the attempt to identify the genes responsible for some medical conditions. In the digital age, in particular of AI, the reduction is from the living as a whole to some processes that can either be described in computation terms (algorithms, programs, hardware) or understood by some so-called intelligent procedure (such as AI deployed for interpreting X-rays or MRI images). Basically, the computational and AI perspectives belong to the deeply rooted view of the human being as a machine (a notion originating from Descartes).

Humans build machines and ended up ascertaining that they themselves are machines. This goes back to the origins of every theology and is the outcome of hypostatizing:

> Those who make an image, all of them are useless,
> And their precious things shall not profit;
> They are their own witnesses;
> They neither see nor know, that they may be ashamed.
> Who would form a god or mold an image
> That profits him nothing?
> ….
> Indeed he makes a god and worships it;
> He makes it a carved image, and falls down to it.
> ….
> He falls down before it and worships it,
> Prays to it and says,
> "Deliver me, for you are my god!"
> ….
> A deceived heart has turned him aside;
> And he cannot deliver his soul,
> Nor say, "Is there not a lie in my right hand?"
> Isaiah 44:9-20 (Guzik 2018)

Within the machine-based theology of medicine, irregular heartbeat (sometimes life threatening) is subjected to the control mechanism of cardiac pacemakers. Currently, they are used successfully for helping patients who need them. Many lives have been saved, and many others are maintained through their use. The first kidney transplants (at beginning, between identical twins) were performed. Heart transplants (more adventuresome than imagined) are performed; artificial hearts (and later, kidneys) are designed and manufactured. Of course, this is not a complete record. Other attempts (e.g., artificial eyes), courageous (if not at times extremely daring), that failed would belong as well in a more comprehensive narration. To celebrate the spectacular successes of the machine-based theology of medicine is not optional. However, to ignore their long-term consequences should mean to disrespect medicine. Justifying the dependencies it creates—some of the happy runners on artificial knees and hips are future invalids, brought to a condition of indefinite reliance on others—means to become accomplice. Likewise for dependence on chemical means—starting with tranquilizers, sedatives, and progressing to pain medications that lead to addiction (sometimes deadly—the current opioid crisis is the example *par excellence*—) to the chemicals that make them effective. Most important: healing as a natural process is quite different from short-term mechanical fixing.

## Is automated medicine still medicine?

After all is said and done, medicine and its associated pharmacological component cannot claim fundamental achievements comparable to those of science and technology. Genetics—so important to medicine as a valid path to healing—is rather the offspring of biology. Making itself dependent upon the science of embodied matter and the causality model of determinism, medicine cannot renew itself and build its own body of knowledge. As an applied endeavor *par excellence*, medicine would not question its own condition as a science and art of healing and furthering life. More recently, under the pressure of the information science model—a data science model, in reality—it has succumbed to the temptation of describing its own goals, means, and methods within an algorithmic approach, often not sure of what an algorithm is. That the meaning of the algorithm escapes the understanding of most practitioners is not a reproach that should offend any of them. Even practitioners and theoreticians of computer science rarely understand what Turing—the mathematician of algorithmic computation—demonstrated: there is no mechanical procedure for proving mathematical theorems. He took Hilbert's and Ackermann's challenge—Can we automate the creation of mathematics? (the famous *Entscheidungsproblem* 1928)—and in answering it in the negative, defined ways to automate the application of mathematics. His machine is not for generating a new mathematics, but for automating mathematics. Wolfram's superb *Mathematica* illustrates the thought (Wolfram 1988). AI comes into the picture because, while claiming to achieve intelligence—which medicine relies on—it actually automates tasks we associate with intelligence. Playing chess is only one example. Interpreting an MRI image belongs to the same category.

Within the obsession with automated data processing, which computers perform better and better, the harness of "precision medicine" was placed on the healthcare economy. Of course, the "complete transformation in therapeutic medicine" predicted for the year 2020 might have to take more time. The "Medical and Societal Consequences of the Human Genome Project" (Collins 1999) are not better than those of the illusory Precision Medicine, sometimes presented under the banner of "personalized medicine," when in reality it is depersonalized to the extreme.

Without any doubt, in the case of some rare conditions, the sequencing of DNA (at huge expense) afforded meaningful clinical evaluations. But not yet necessarily better treatment. "Those who have followed the gene-therapy field over the decades may be weary of forward-looking positive statements" (High & Roncarolo 2019). The complete costs of genome sequencing that could be used in routine healthcare has decreased from \$20-25 million in 2006 to \$25-30 thousand, plus costs for analysis (Schwarze et al 2019). Assuming that each person could have the genome map associated with his or her data, we would still not know more about the Why? of health (or lack hereof) and how to address its maintenance. The data do not carry meaning with them, but rather become meaningful once they are associated with the context of medical concern. Furthermore, epigenetic factors have also a say in the matter of how health can be compromised or, alternatively, regained and maintained.

The threshold of G-complexity is merciless: hypertension, diabetes, quite a number of cancers, not to mention depression, obesity, and extreme emotional states are the expression of a very large number of genetic variations and of epigenetic factors. It is not a matter of how large the data sets are; it is the intrinsic condition of the living: the impossibility to fully and consistently describe the process. Moreover, no reduction, typical of deterministic systems (where it works), is possible. Genetic expression at the holistic level, including the microbiome, stands in no relation to subsets associated with medical conditions. Whether scientists are aware of it or not, genetic expression takes place in a context. Assuming that the processing of genetic data can be automated using an algorithmic procedure, this will not compensate for the fact that algorithmic processing is context dependent. Moreover, genetic processes do not align with biological processes—epigenetic factors, for instance. Therefore, actionable knowledge cannot be derived from genetics to the benefit of the broader biological condition. Gene expression is anticipatory. Cells are in preparation of what follows, not in reaction to what was. The DNA testing for breast cancer, which is more than what mammograms provide, returns at best "polygenic risk scores," which are rather similar to the risk scores for earthquakes and tornadoes. Probabilities are quite adequate for describing the non-living, but they are not optimal for the practice of medicine.

Similar considerations apply to pharmacogenomics. The actions of drugs is always risky: the so-called side effects prove to be more dangerous than the alleviated health condition. There has been some success in revealing unintended damage or actual ineffectiveness, but all in all, pharmacogenomics as a deterministic procedure does not justify the immense investment (human effort, money, testing, validation) it has required so far. That AI applications are conceived with the purpose of better medicating patients does not make medication neither more adequate nor less dangerous (through the side effects).

Targeted therapy, for which AI methods of fitting a curve to data would be deployed, remains no more than yesterday's promise. Without going into further details, let us take note of the fact that preventive medicine—a very honorable goal that everyone can only root for—has not made progress. Genetic testing, whether AI-based or conventional, does not de-confound factors of illnesses. Worse, genetic knowledge, as dubious as it is in its reductionist expression, does little if anything to affect behavior. Gene therapy, which was intended to support targeted therapy, has not yet fared better. Epigenetic considerations could make a difference in both addressing behavior and a new perspective of curing illness (e.g., cancer) if its recent progress will translate into some operable treatment paths.

## Turning the doctor into a machine

In the process of socializing by decree insufficiently tested scientific means and methods, medical practitioners were transformed into cogs in the vast machinery that medicine has become. With few exceptions, mainly in alternative medicine practice (sometimes undermined by charlatans, taking advantage of this less regulated field), doctors welcome the patient like mechanics welcome cars needing repair. Computers, present not for assisting, but rather for bureaucratic reasons (HER/Electronic Health Record compliance), fully obstruct views of the patient. Typing (even in the age of Cortana, Siri, and Google Assistant) takes the best of the doctor's effort. Indeed, physicians have access to data and can recite whatever is standardized. They can retrieve useful information as needed from large databases (the FHIR: Fast Health Interoperability Resources decree). Each test intended to help in the diagnostic comes interpreted. This pre-automated medicine made Zittrain describe how AI, like asbestos, infiltrated medical practice, frequently without the physicians' knowledge. It practically eliminates patient-physician interaction, not as their choice, but as a result of bureaucratic rules. Self-service in healthcare is around the corner—like automatic teller machines and other online transactions that eliminated bank tellers and quite a number of human service providers. This is the equivalent of the McDonald's method—production line for fast foods—or tax returns using Turbo Tax (one app for all!). Serious MDs brought up these examples in the context of "self-service medicine" (Asch, Nicholson, Berger 2019) while paradoxically arguing for support of this model. The Patient Portals, to which patients are corralled within the current political understanding of healthcare are only the forerunners of yet another revolution, i.e., when Deep Medicine, takes over, and all of a sudden—miracles happen—doctors will find time for their patients because of AI. It turns out that even the proponents realize that they might have gotten it wrong (Muse & Topol 2019).

## Can AI (really) make healthcare human again?

Infomercials aside, AI is omnipresent in medical journals and at professional conferences. "How Artificial Intelligence Can Make Healthcare Human Again" (the subtitle of the book *Deep Medicine*) is representative of the tenor of the discussion on the matter. For an informed conversation of the subject, it is necessary to understand not only what healthcare is, but also what it takes to qualify as a valid perspective for the practice of medicine. The very large body of knowledge accumulated in medicine constitutes the foundation of medical education and for ongoing training. As already made clear, the foundation involves knowledge of science and technology in the broadest sense in which these words are used. Nobody expects physicians to be experts in plate tectonics, earthquakes, or volcanic activity. But physics, chemistry, and data processing are imparted to those seeking accreditation because they are relevant to dynamics of the living. The right to work in healthcare—the golden cage, as some see it—is associated with all the responsibilities the profession implies

In addition to knowledge, which can be formalized—there are knowledge repositories available in the form of medical decision-trees, which can be accessed against payment—there are skills to be acquired and maintained. No doubt, the new means and methods of knowledge representation and management associated with computation will help those in medical care effectively navigate the rapidly expanding acquired expertise characteristic of this particular form of human activity. By the nature of the profession, physicians exchange information and experiences because the outcome, in ideal form, is life, not a competitive edge in adjudicating profit or monetizing some new ways to help patients.

Before even entertaining the straight-forward question of what can be expected from a generalized deployment of AI capabilities, it makes sense to identify how human intelligence partakes in the discharge of the duties of healthcare professionals. This applies to everyone, from those maintaining sanitary requirements (in offices, clinics, and hospitals) to nurses, assistants, physicians, and, of course, administrators of healthcare businesses (the highest paid segment within the medical complex), including insurance companies.

For the physician's cognitive profile, it is useful to know that the IQ of medical doctors is at the level of professionals in the natural sciences (e.g., physics, mathematics, biology), of lawyers, and of college professors. Even those doubting the relevance of the IQ metrics would accept that the span between 110 and 130 and above testifies to a rigorous selection process. The ten years it takes to attain the degree and the following accreditation, are also testimony to a sense of assumed and exercised sense of responsibility. It also explains their analytic skills, the ability to interpret data presented in alphanumeric form or in visual representations, their tendency to seek associations and correlations, to make inferences, and to question themselves. It is an activity that implies professional ethics—pathogenesis and ethos should be seen in their connection—commitment, and communication skills. For all practical purposes, doctors are small business managers, and not rarely engaged in research, as well as in publishing their findings and experiences. Being science, technology at work, and art at the same time, medicine is basically grounded in the interaction between healthcare provider (physical therapist, nurse, surgeon, physician, etc.) and those who seek qualified help in maintaining or regaining health. A concept that frequently comes up, especially in the context of contemplating automated medical procedures, is emotional intelligence. In particular, empathy—to feel for the other, and to feel what the other is feeling—is emphasized.

On this note, a first observation begs for our attention: Medical education imparts knowledge, but is also supposed to make students aware of the role that empathy plays in the day-to-day healthcare delivery. Research has shown that during medical education empathy actually deteriorates (Neumann et al 2011). Despite the evidence, the Why? of this situation is almost never addressed. It is the result of the machine model, or of conveyer-belt medicine (Elia & Aprà 2019) not only adopted in the practice of medicine, but also driving medical education.

Mechanics and assembly line workers do not need empathy in order to fix or make cars, airplanes, or boats. For medicine, the loss of empathy results in the lower effectiveness of treatment. Empathy supports the effort to engage the patient. This observation will serve us well when we examine the expectation that AI will free the physician from some tasks, and thus give them the opportunity to better express empathy.

All these background factors—empathy stands out—are essential if the task of evaluating how AI will affect medicine is taken seriously. In a different context, I argued that while the qualifier "artificial" in AI is beyond controversy, intelligence is not (Nadin 2018b). To automate activities in which intelligence, in some of its many forms of expression, is involved is not the same as making intelligence available. Medicine is a good example of this. Those practicing it are confronted with the definitory aspect of intelligence: to understand *before* you act, not necessarily to act in a manner that afterwards *seems* intelligent (Nadin 2018c). This is what a physician does: associating symptoms with possible causes against the background of the patient's personal narration, i.e., the timeline of events from birth to the moment when the consultation takes place. Understanding is context dependent and predates the action. For this understanding to arise—intelligence is a process—there are quantified aspects (measurements) to be considered; there are also qualitative assessments to be made; and there is the empathy. In a description that intentionally goes to the extreme, empathy means that a doctor experiences what the patient is going through. The pain of the others becomes the doctor's pain. They die with those dying in their hands. This is not poetry. Let us recall the mirror neuron cognitive model (Gallese & Fadiga 1996; Rizzolati & Craighero 2004;) and its experimental evidence. The configuration of neurons in a learning situation imitates that of the teaching person. Other scientists submitted to the scientific community and to medical practitioners proof that the mirror neuron system underlies empathy (Preston & de Waal 2002; Decety 2006; Jabbi & Keysers 2008).

Based on this observation alone, not on sentimental descriptions of the role of emotions, we understand why the burnout among those who work in medical care is higher than in other segments of the population. We also understand why the suicide rate exemplifies the existential nature of the activity of physicians, regardless whether they view their patients as machines in need of fixing, or as living persons in need of individualized interventions of genetic or epigenetic nature (self-repair, for instance). Diagnostic procedures, some well-defined (as in differential diagnostics) involve the empathy component. Healing is often (but not always) self-healing, in the reality of the integrated human being in which each cell is literally involved. To ascertain that empathy will again be made possible when AI takes care of tasks that can be automated is indicative of machine theology: we made them, they can replace us, provided that we join the “church” (or the cult, as deep learning has become).

On a daily basis, the industry trumpets yet another success in automating the evaluation of radiology images, of MRSs, of CT scans. This is worthy of acknowledgement because better descriptions (through measurement or visualization) are helpful. But it is a self-deceiving trend. The obsession with measurement, with more data, created the bottleneck in the first place—and made the radiologist the best-paid medical practitioner. Now deep learning is supposed to do the same: evaluate data, but at a fraction of the cost—that is what automation means—in disregard of the disconnect between the living who is undergoing change, and the non-living object that is monitored in order to avoid breakdowns.

To associate personal history data (each patient brings a history, from birth to the current condition, with himself or herself) and imaging data—the present visualized—is a goal worth pursuing. To assume that it can be attained on account of AI (in the classic definition) or deep learning is more a wish that a conclusion based on understanding what such technologies can achieve. The anamnesis (the history) is an interpreted timeline: it invites a new evaluation, i.e., considerations expressed at the semantic and pragmatic levels of the representation. No computation based on the Turing machine is appropriate to the task. It takes place only at the level of the syntax. The expectation that the newly generated visual image (e.g., a fracture, a blood clot) will be delivered together with its interpretation is to imply that "There is a god in the machine," since context escapes mathematical description—without which no computation is possible. A mathematics that interprets itself is a nice idea for science fiction, but otherwise, it is symptomatic of ignorance. Moreover, what makes the new optimism (some might define it as an "expression of ignorance') even less acceptable is a limitation that goes deeper: each living being is unique. There is no way to generalize within the idiographic—the open-ended space of uniqueness. To understand that the normal can be bad for one person, or the abnormal good for another, only suggests the implications of the uniqueness. To deal not with quantities, but with meanings—not within the functions of AI or algorithmic computation—is yet another expectation waiting to be acknowledged. It might well be that AI predicts death better than physicians can (Weng et al 2019), but to fight death, and often succeed against the odds, is quite different in meaning from death by numbers.

## How well do physicians and patients understand what AI is?

Understanding the science behind AI is the prelude to the decision of whether it makes sense to invest in learning how to use it. Science, and in particular medicine, is driven by optimism. No science worthy of its name starts off with “It is impossible.” No medical assessment ends in less than optimistic terms, even in cases of the still incurable. All those understandings of reality concretized in the knowledge of hydraulics, pneumatics, optics, electricity, combustion, which medicine adopted as acceptable descriptions of the human being, originated in the human mind. They were subject to being understood, and were tested in reality. For a time, knowledge seemed independent of the originator. Like: “It is real that intestines are like irrigation canals,” therefore if you experience pain because they get clogged, the physician has to help open them up. Or, to come to our time: the brain consists of neurons, therefore our mathematical description of the neuron applies to our brain and explains how it works. From here to the theology of neuronal networks, of deep and deeper learning, the jump is pretty dangerous if not enough knowledge is used. The optimism, by no means uncalled for, at least in its spirit if not in its literal meaning, of a Hinton (already mentioned) should be celebrated, but not automatically generalized to medicine, which is a different knowledge domain. The “new Descartes,” not of mechanical clocks (the machines of the 16$^{th}$ century), but of artificial neural networks (Hinton 2018), might realize, as a potential patient (and who is not a potential patient?) the difference between the living (and its intelligence) and the machines supposed to help doctors and patients.

Here we should take note of an interesting parallel: What distinguishes the training of neural networks (i.e., machine learning) and the training of medical professionals. Deep learning is driven by a huge amount of data and takes place at a cost of impressive energy use. The outcome is the convergence on a specific diagnostic. The learning for becoming a physician is focused on making inferences possible: from a specific case to the generality (sometimes a spectrum) of various conditions. The open-endedness of medical conditions is met by the adaptive nature of the physicians. Machines can be extremely precise, but do not evolve and cannot capture the dynamics of life. The energy expense for training a physician stands in no relation to that used in machine learning.

Neither Thomas Bayes of the Problem in the Doctrine of Chances (1763); nor Adrien-Marie Legendre with his “least squares method” (Méthode des moindres carrés, 1805); nor Pierre-Simon Laplace on probabilities (1814), nor Markov (1906) analyzing a poem without realizing that the technique would become a powerful tool in the age of data processing); and probably not even the polymath Alan Turing (1950), had an inkling that healthcare would eventually be affected by their descriptions. “The Beginnings of Artificial Intelligence in Medicine” (Kulikowski 2019)—a very thorough report—reflects the understanding of healthcare as a heterogenous knowledge domain (Kulikowski was part of the beginnings) associated with skilled performance, some subject to automation.

Within Artificial Intelligence in Medicine (AIM), physicians are identified as problem-solvers, using *ad hoc* heuristics. Their reasoning, in this view, is based on pattern recognition. Physicians’ decisions are the outcome of clinical data processing and of interpretations often tested within the feedback mechanism of the patient-doctor interaction: “How does this treatment work?” “How does this dietary prescription affect my condition?” etc. The boom and bust (called “AI Winter”) in AI is reflected in the waves of optimism (excessive at times) that followed each disappointment, especially in the practice of medicine.

## The role of ontology

If there is one undeterred development, it is the development of ontologies, the common medical vocabulary, in some machine interpretable definitions. The National Library of Medicine (NLM) in the USA recognized early on (in the 1960s) that the promises of data processing would come to fruition only if a mapping from the language of healthcare professionals to computation could be made available. Chances were close to nil that the mathematicians, logicians, and computer scientists who dedicated themselves to AI were also competent in medicine. Moreover, the large repositories of knowledge (biomedical literature, vocabularies, encyclopedias) could not be implicated in the effort without providing access in some machine language. The WordNet (1980, by George Miller at Princeton University) was an example of how to organize language in machine-readable logic. The foundation for AIM was provided by Medical Literature Analysis and Retrieval System (MED LARS) and its successors (MEDLINE), as well as the associated search engine (PubMed). Physicians and all those involved in medical care developed a language appropriate to that aspect of reality—health—that made their activity necessary. To describe the patient in a language that supports easy retrieval, reuse, sharing, and eventually actions is an accomplishment for which medicine never gets enough credit. Ontologies describe what is. Celsus (25 BCE-50 CE) referred to color (*rubor*), heat (*calor*), shape (swelling, which is *tumor*), and pain (*dolor*) in order to produce the ontological equivalent of inflammation. The challenge is not to translate the early description into machine language, but to provide means to translate all the ambiguities into its language. Intelligence means to understand, which is a mapping from a description—words, images, sounds, etc.—to some action. Data describe the formal aspects of the real: numbers identifying what kind of red, how high the temperature, how fast the heartbeat, etc. Dictionary definitions map to common use: how patients and physicians describe an inflammation. Meaning is at work in respect to action (the targeted therapy): what to do to address the processes leading to an inflammation. And, no less important: how to make sure that "fixing" something does not lead to "breaking" something else.

Putting in the proper light the language involved in "reading" the symptoms, in issuing a diagnostic, and in formulating a course of action (the desired remedy) is relevant because in the final analysis, AI, in both its symbolic and statistical embodiments, is driven by ontologies. ImageNet (Li Fei Fei & Olga Russakovsky in 2007), the visual parallel to WordNet, made available millions of annotated images, organized in a large-scale Hierarchy Image Database (HID). The AlexNet (Krizhevsky, Sutskever, Hinton in 2012) and the OpenImage (Google in 2016) helped in the integration of word and image which, at least for medicine, was a necessary condition. The dictionary defines an inflammation as a local response to cellular injury. It defines its symptoms as capillary dilation, leukocytic infiltration, swelling. Ontologies of inflammation—here used as an example (inspired by Pisanelli 2004)—are more than dictionary descriptions. They provide data actionable upon. There are agents (physical, chemical, biological) that cause an injury. Affected blood vessels and adjacent tissue change their state due to the stimulation. The removal of the injurious agent and the stimulation of repair processes are a course of action that the physician adopts in awareness that the meaning of the inflammation is not the same as the data describing it. The variety of attributes—amount of infiltrated substance causing the injury, type of lesion (superficial, deep, etc.), duration, and similar—of the process is evaluated in the context of the patient's condition. A compromised immune system, suggests a different path of medical intervention than the immune system of a healthy patient. The Unified Medical Language System (UMLS), which the National Library of Medicine initiated, is a repository that helps in defining clinical guidelines. Information processing supports retrieval, either by the human physician or by some intelligent agent. Without any doubt, statistical inferences, on whose basis deep learning comes about, and quantitative meta-analysis have contributed a lot to medical knowledge acquisition and accumulation. They also made AI in medicine a reality, regardless of anyone's skepticism. Imaging, as a diagnostic tool, is an area where the interpreting radiologist and AI—deep learning trained on huge data sets—came to compete with each other. Automating malaria detection by making possible the assessment of malaria parasites in a blood sample is based not only on processing vast amounts of data, but also of having the proper ontology in place. What a digital camera attached to a microscope "sees" and expresses in actionable data is defined through the appropriate ontology.

Ontologies are behind natural language processing in yet another way, through which the syntax-driven computer is associated to dictionary-defined meanings of words. Suicide, a major cause of death in the USA, associated with mental health challenges, cultural aspects, and socioeconomic conditions, is rarely (if ever) a spontaneous action. Within every suicidal person there is a narrative of behavior and there are specific forms of expression. AI-driven "scanners," combing through social media messages (*scraping*, as it is called), but also through health records, could identify suicide risk. The fact that privacy is at stake, and the decision of what weighs more—prevention or risk—cannot be ignored. This applies to all aspects of monitoring: wearable devices, communication (digital media, such as e-mail, digital telephony, messaging, etc.).

## Medicalizing the healthy

Of course, AI can fully automate the imbecilic bureaucratic overhead of regulations (EHR and even part of FHIR) and free the physician from the tasks of typing or voice inputting to recording devices. But even for this worthwhile task, the dangers of abandoning privacy, which medicine has so far protected, are real. The mobile wearable devices are a promise that makes those in the AI business salivate (Sim 2019). Sensors continue to diversify. Their sensitivity is increasing. Access to data is instantaneous. Connected to networks of all kinds and to the cloud, such devices are the new promise of good sleep, good nutrition, good exercise, entertainment escaping loneliness—you name it, they do everything. But society learned that the dangers associated with their use are often as great as the opportunities.

Medicine should begin with prevention, and not take the path of measuring more and more. This very simple premise can mean many things, among them, the extreme: measure everything every time. Lisa V. Hamill's tweet on the matter went viral:

> The USA has been accused of an over-focus on tests and drugs, practicing expensive care with less than stellar results. Now we turn to "medicating good health?!" You can't be "healthy" without monitors and tests? You need 8 monitors on your wrist to tell you to exercise/lose wt?

There is ample evidence for associating the obsession with all kinds of devices (from the innocuous Fitbit™ and Applewatch™ to sensors that monitor sleep, eating patterns, sexual activity, for instance) with possible negative effects. Start with the psychosis—e.g., blood pressure does not improve through continuous checking, but becomes an obsession—and continue to a large number of interventions not really meaningful since they could actually undermine a person's condition. John Mandrola (2019) reports on a case in which some abnormality, usually ignored, prompted an intervention that led to a stroke in a retired but still functioning farmer, thus forcing him into a nursing facility. Interestingly enough, even the proponents of phenotyping as the backbone of Deep Medicine realize that more data is a "kitchen sink" obsession that can backfire. "Creation of revenue," as Muse and Topol (2019) call it, is a delicate way of describing medicine driven by greed. Ivan Illich (1974), critical not only of the education system, but also of medicine, uses the word *iatrogenesis* to describe clinical harm from excessive screening—only because we can screen and can carry a device on us. Digital implants are tampered with; malice was injected into medical imagery. Medical data breaches and interruptions in medical services are in line with acts that not only cost money, but also undermine the social fabric. Insurance fraud, sabotage research, political malice, and media manipulation are not the same as injecting a lung cancer into a CT scan. A generative adversarial network (GAN), which is a neural network within which the distinction between real and fake samples serves as a learning process, could be used not only to distinguish between a cat and a dog, or between healthy and unhealthy cells, but also for malicious purposes (Mirsky et al 2019). It can affect picture archiving and the communication systems (PACS) that manage CT and MRI scanner data. All this sounds more like escalation of various societal conflicts than progress in healthcare

It is a sign of responsibility that there are voices warning against the consequences of creating dependencies, some of which can lead to harm. To repeat: medicine and ethics cannot be separated—pathogenesis and ethos are co-substantial. On the other hand, the amount of dedication and enthusiasm of those who examine the new opportunities is encouraging. New ideas come to the fore; experiments are conceived and carried out; the optimism inherent in science extends into the medicine of the time of AI and of many other scientific and technological innovations.

## Deep Medicine revisited

The subtitle of the book *Deep Medicine* provides an image of what the medicine of our time (and of the future) might become. "How Artificial Intelligence Can Make Healthcare Human Again," is an issue that Eric Topol (2019), himself one of the most influential authors addressing the broad public tries to answer. Topol, a distinguished cardiologist (still practicing one day a week) and an early adopter of digital technology in medical care, proudly discloses his involvement with companies producing such technologies. He follows the path of those scientists (such as Marvin Minsky, Craig Venter, Stephen Hawking, Carl Sagan) who became the public face of new fields of human inquiry. He joins physicians who cover medicine for TV networks and major media outlets. In ten years of sharing information via Twitter, Topol single-handedly produced over 18,000 messages (the so-called "tweets"). He is informed about what is taking place, and he is passionate about a human-face medicine. With the ill-conceived notion of Deep Medicine—following in the footsteps of calling the shallow techniques for processing statistical data "Deep Learning"—Topol submits a model of medicine that can be criticized but not ignored. His rather vivid prose (with good examples from his practice) is meant to describe a three-prong approach:

1) A complete description of the individual. In his view, deep phenotyping means a full mapping—from the data pertinent to one's medical condition to social, behavioral, etc. records. Of course, DNA, RNA, proteins, microbiome, etc. are part of this mapping.
2) Deep learning—including visual patterns, processing of data associated with symptoms, even nutrition patterns. Medical care shifts from direct contact between the physician and patient to a virtual interaction.
3) Deep empathy, which can result from freeing medical caregivers from anything that interferes between them and their patients.

Reviewed by everyone, in almost every publication (it is worth having a good literary agent), the book is significant for assessing the level of understanding that both medicine and the new technologies have summoned. The goal: to save medical care from a condition of subservience to regulation, economic pressure, and technological dependence. Given the fact that everyone's life is, in one way or another, sooner or later, affected by the state of medical care, the new paradigm deserves full attention—no less than AI itself or, for that matter, computation. (One previous paradigm was called "Digital Health.")

Practitioners and academics debate whether "…artificial intelligence makes doctors obsolete?" (Goldhahn et al 2018), not so much because they understand the implications of the question, but rather because they are under pressure—from everyone, patients included. The record of achievements—in image analysis (X-rays, retina scans, MRI, etc.), in genetic assessment (based on genome scans), in clinical decision support (such predicting septic shock), virtual nursing (keep patient under remote observation via the internet), robotic surgery (laser eye surgery is in the lead), and more—is as impressive as the record of failures. IBM's Watson Health Division is the first example that comes to mind. At this juncture, we could reference thousands of reports on achievements, but no less on failures. However, their meaning will escape usefulness. We will not know, from the reporting, to which extent they are accidental or reflect a new horizon. In order to understand what such performance means, we have to understand the epistemological background against which we can evaluate them.

## Back to Shannon – the question of entropy

More important from the perspective of medicine and its exposure to computation and AI is yet another contribution by Shannon: the realization that data transmission is affected by what the second law of thermodynamics describes. As a matter of fact, entropy, which characterizes the disorder of a system (to use a simplified description), affects data transmission. With this additional aspect in mind, it becomes evident that data processing within a relay circuit, an integrated transistor circuit, in a liquid (the test tube DNA computation of Leonard Adleman), in an artificial muscle, in any medium is subject to limitations resulting from the entropy of the medium used. Quantum computation—the new frontier in computation—forced those who are trying to achieve the performance increase it promises to invest a lot in securing an environment in which entropy can be controlled (Nadin 2014).

Since the living is negentropic—its entropy does not increase as in physical systems—the question is whether computation, i.e., the automated mathematics it facilitates, can properly describe living processes. By extension—since AI, regardless of its kind, is achieved through computation—it is fair to question the extent to which it can do more than assist a physician in addressing the state of a patient by providing the benefits of automated data processing. But which data? All that there is? Or actually the significant data that a good doctor identifies?

Like everything ever invented, the computer is a deterministic device. For everything of deterministic nature in the dynamics of life, computation-based means and methods are useful. Just for the sake of providing some examples: extreme precision surgery, guided by large sets of numerical descriptions, benefits from computer-driven surgery tools. The robots deployed for such types of interventions have reached a performance level that cannot be matched even by the most experienced surgeons. Things are not so clear-cut when it comes to tasks that by their nature qualify as non-deterministic. Illness, as the medical community realized over many centuries, is an ill-defined problem. It does not suffice to measure more and more and to process higher and higher volumes of data in order to assess a patient's condition. Not to say in order to define a course of action that might cure the patient, or at least provide means to alleviate many aspects of illness. Topol's book, the "bible" of this moment, ignores all these questions.

If we want to accelerate the process of transforming the human being into a machine, AI (and computation) is the way to go. If, alternatively, we want to free medicine from the condition of being a repair shop for human beings, we need to change the perspective of medicine. Topol does not recognize the need for a new perspective. We argued in favor of anticipation-grounded medicine as the necessary alternative (Nadin 2018c) (Fig. 3).

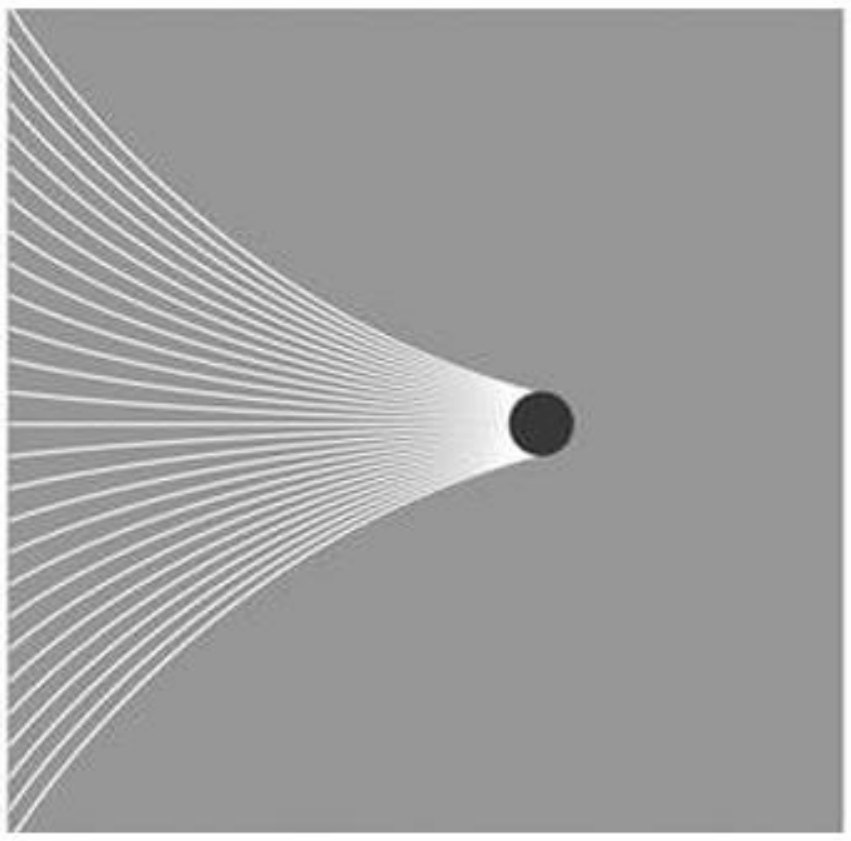

Fig. 4. Anticipation-grounded medicine corresponds to the anticipatory nature of living processes.

Indeed, medicine, and those working hard to help healthcare practitioners, need to rediscover the living as having a condition different from the non-living. Of not being a machine! AI and computation should be considered for extreme situations in which our knowledge of the living is still so rudimentary that we have to make use of physics and chemistry, instead of involving biological means for maintaining health or for healing.

**Acknowledgments**

Many practicing physicians, to whom I wish to express gratitude for their tolerance of someone who questioned their profession, educated me without turning me into a practicing doctor. They read this paper and offered very serious feedback. No agreement was actually reached on any of the major value judgments made in this paper regarding the danger of a new *theology of medicine.*

In particular, I wish to express gratitude to Jean-Paul Pianta, Matthew Goldberg, Thomas O. Staiger, and Oleg Kubryak.

The Ontolog-Forum, in particular through Azamat Abdoullaev and John F. Sowa, educated me in matters of medical ontology. Stuart Kauffman and Cristian S. Calude brought to my attention the work of Giuseppe Longo.

Funding for the research on which this paper is based was provided by the antÈ-Institute for Research in Anticipatory Systems.

Dr. Asma Naz was, as usual, prepared to help me make ideas in this text more accessible to a larger audience. Elvira Nadin could claim at least co-authorship by challenging almost every hypothesis herewith formulated and eventually tested in our lab.